\documentclass[11pt]{article}
\usepackage{latexsym}
\usepackage{graphicx}
\begin{document}

\title{Weighted networks of scientific communication: the measurement and topological role of weight}
\author{Menghui Li$^1$, Ying Fan$^1$, Jiawei Chen$^1$, Liang Gao$^1$, \\ Zengru Di$^1$, Jinshan Wu$^{2}$\footnote{Email Address: jinshanw@physics.ubc.ca}\\
\\ 1. Department of Systems Science, School of Management,\\
Beijing Normal University, Beijing 100875, P.R.China
\\2. Department of Physics \& Astronomy, University of British Columbia, \\ Vancouver, B.C. Canada, V6T 1Z1}

\maketitle

\begin{abstract}
In order to take the weight of connection into consideration and
to find a natural measurement of weight, we have collected papers
in Econophysics and constructed a network of scientific
communication to integrate idea transportation among
econophysicists by collaboration, citation and personal
discussion. Some basic statistics such as weight per degree are
discussed in \cite{fan}. In this paper, by including the papers
published recently, further statistical results for the network
are reported. Clustering coefficient of weighted network is
introduced and empirically studied in this network. We also
compare the typical statistics on this network under different
weight assignments, including random and inverse weight. The
conclusion from weight-redistributed network is helpful to the
investigation of the topological role of weight.
\end{abstract}


Pacs: 89.75.Hc, 89.65.-s, 89.70.+c, 01.30.-y

\newpage

\section{Introduction}
Recently many researchers in different fields use the topological
properties and evolutionary process of complex networks to
describe the relationship and collective behavior in their own
fields\cite{albrev, evol}. This methodology, which is so called
network analysis, often leads to discoveries. Also new analysis
methods and new topology properties are proposed by this approach.
A network is a set of vertex and a set of edges which represent
the relationship between any two vertices. Just because of its
simplicity of this description, network can be used in so many
different subjects (see \cite{albrev} and its references),
including linguistics, collaboration of movie actors and
scientists, human sexual contacts, disease propagation and
controls, community structures, information networks, and food
web.

However, a single line representing the existence of the relation
will be a limitation when it is used to describe relations having
more than one level. For instance, in the network of scientists,
both collaboration and citation are the ways of idea
transportation but with different contributions. When we analyze
this transportation as a whole, we have to use different weight to
measure these different contributions. Also, even for the same
level interaction, such as collaboration, not only the existence
of connection but the times of collaboration is a valuable
information. So to fully characterize the interactions in real
networks, weight of links should be take into account. In fact,
there are already many works on weighted networks, including
empirical studies and evolutionary models
\cite{effweight,weight,rousseau,foodbif}.

The way to measure the weight for weighted networks has been
introduced differently in several types by some authors. First
type, transfer some quantities in non-weighted network into the
weight of edge. In \cite{Macdonald}, the weight of an edge is
measured by the point degree $k_i$ and $k_j$(e.g. $w_{ij}=k_ik_j$)
of its two ends. Second type, in some networks, typically nature
measurement of weight is already given by the phenomena and event
investigated by the network. In the scientific collaboration
network, the times of co-authorship are registered as the weight
of link\cite{newman}. In \cite{Barrat1}, in the case of the WAN
the weight $w_{ij}$ of an edge linking airports $i$ and $j$
represents the number of available seats in flights between these
two airports. In \cite{Bagler}, the weight $w_{ij}$ stands for the
the total number of flights per week from airport $i$ to airport
$j$. The third type is in works about modelling weighted networks.
Some prior weights are introduced\cite{weight}. In \cite{Park},
the weight $w_{ij}$ of a link $l_{ij}$ connecting a pair of nodes
($i$ and $j$) is defined as $w_{ij} =(w_i+w_j)/2$, where $w_i$ is
defined as $i$ node's assigned number (from 1 to $N$) divided by
$N$. In \cite{Antal,Goh2}, the weight $w$ is assigned to the link
when it is created, which is drawn from a certain distribution
$\rho(w)$. In fact, the first type of weight description should be
regarded as an approach of non-weighted networks. It is helpful to
discuss new properties of the non-weighted networks but without
taking any more information than the non-weighted networks about
the real interactions. In the second type, which is a very large
class of the weighted networks, typical measurement of weight is
already given by the phenomena. The investigation of such network
focuses mainly on how to define and discover the topological
character of the networks. In the last type, from the viewpoint of
empirical study, we never know such models already acquire the
real structure of weighted network or not. In fact, giving some
hints on modelling weighted network is also a part of the goals of
our empirical investigation. The empirical study of weighted
network without a naturally given definition of weight, is
especially valuable to answer following questions, such as how to
define a well-behavior weight, and how to extract structural
information from networks, and what's the role of weight according
to its effects on the structure of the network.

In our work, we apply the general approach of weighted network
analysis onto network-style phenomena without given measurement of
weight. There are lots of such kind of networks. For example, in
our case, we try to construct and reveal the structure of the
network behind the transportation of ideas in scientific
community. Actually the scientific collaboration has already
become an interesting subject for network
research\cite{redner,scievo}. But in this paper, how close two
scientists related in our network and how easily the idea
transferred between them, are the phenomena we are interested in
this network. This is similar to the situation that one want to
construct and reveal the structure of network of underground
railroad by the information about traffic, the passengers coming
in and out at stations, without a map of the subway. Therefore,
both the existence and the times of coauthoring (or citation,
acknowledgement) are important for the network construction. And
the times, for sure, implies some information about ``how close
and how easily" in the sense given above.

In order to extract relationship information from the times of
interactions, a tanh function is used to convert the times into
weight, and all the weights from coauthor, citation and
acknowledgement are combined into a single weight of every edge.
Tanh function starts from $tanh(0)=0$, and increases up to 1 when
variable is large enough. The times of the event, in our network,
is a cumulative number. Intuitively, the more times, the closer is
the relationship, and the less contribution that one new event can
provide to the relationship. That means the contribution of a new
event to the relationship should decrease on marginal. The reason
of such a saturation effect is that, what we want to analysis is
the relationship of ``how closely and how easily'' , not the
events of the transportation, although we have to start with the
events and extract information from them. With the subway analogy,
the railroad network of ``how wide is the road between any two
stations'' is our object, not the traffic itself, although the
only information we can make use of is about the traffic. Because
of the same reason we incorporate the three weights into a single
weight. In the sense of idea transportation, they provide the same
kind of information about ``how closely and how easily'', only
with different contributions. Now, the next problem will be how to
measure them differently by their deserved contribution? Frankly,
we have no principal way to measure the ``deserved" contribution.
The thumbrule here is the ratio of total times of the three
events, 7:2:1, is used for their relative contributions. We have
tried to reveal the effect of different relative coefficients. But
the topological quantities and their distributions we have done
now is not enough to describe such effect. It seems the effect of
different coefficients can only be shown by some new topological
quantities. The figure \ref{wnonw} hints that in order to reveal
such effect, we have to come to the correlation analysis.

In reference \cite{fan}, we have constructed such a weighted
network of idea transportation between scientists in Econophysics,
an active field oncoming recently \cite{ecophys1,ecophys2}. Basic
statistics have been presented, including the weight per degree.
In this paper, we collected most papers till July 2004 in
Econophysics, and constructed this networks as a sample of
weighted networks. Now we ask the questions: first, whether the
distribution and property of the basic statistics change after the
one year development; second, whether the way to measure the
weight is significant for the structure of network and what's the
effect on the structure of network if the weights on the edges are
redistributed; third, what are the definitions and properties of
more quantities such as Cluster Coefficient. The matching pattern
in directed and weighted networks, the robustness of weighted
networks and the topological property of weight will be discussed
in later papers. The second part, the effect on network structure
by changing the matching pattern between weights and edges, plays
an important role in this paper. Because we think this
investigation reveals the topological role of weight: does the
weight affect the network significantly, and for a vertex, is
there any inherent relation between its weight and is status in
the network? However, in fact, these questions are not fully
answered in this paper yet. In this work, we randomize the
relation between weights and edges with the similar idea of
randomizing the connection under fixed number of edges in WS
model\cite{swn}. And we think this approach partially realized the
idea about investigating topological role of weight.

Just because the Econophysics is hot in both Finance and
Statistical Physics, our work will be of interest to
econophysicists for another reason: it's about their works, and it
represents the idea transportation between them.

\section{Measurement of Weight and Basic Statistical Results}

Recently more and more researchers in economics take up
Statistical Physics to explore the dynamical and statistical
properties of financial data, including time series of stock
prices, exchange rate, and the size of
organizations\cite{sutton,zipf}. Meanwhile many physicists from
Statistical Physics and Complexity turn to working in finance, as
an important and copies research subject.

To investigate the development of such a new subfield is an
interesting work itself. In our previous paper \cite{fan}, we have
introduced the work of paper collection and the construction of
the scientific communication network. Concentrating on main topics
of Econophysics, we collected papers from the corresponding web
sites. The basic statistical results of the network was given in
\cite{fan}. It was constructed by papers published from 1992 to
4/30/2003, including 662 papers and totaly 556 authors. After
publishing our first paper on this research, we keep tracing the
development of Econophysics and enlarge our database in time. In
this paper, we will give the basic results for the network
includes 808 papers and totaly 819 authors from 1992 to 7/30/2004.

Because the weight is a crucial factor in our network analysis,
here we introduce again the measurement of weight in the network.
Based on the data set, we extracted the times of three relations
between every two scientists to form a file of data recorded as
`$S_{1}$ $S_{2}$ $x$ $y$ $z$', which means author $S_{1}$ has
collaborated with author $S_{2}$ `$x$' times, cited `$y$' times of
$S_{2}$' papers and thanked $S_{2}$ `$z$' times in all $S_{1}$'s
acknowledgement. One can regard this record as data of three
different networks, but from the idea of transportation and
development of this field, it's better to integrate all these
relations into a single one by the weight of connection. Here we
have to mention that in order to keep our data set closed, we only
count the cited papers that have been collected by our data set
and just select the people in acknowledgement which are authors in
our data set.

We convert the times to weight by
\begin{equation}
w_{ij}=\sum_{\mu}w^{\mu}_{ij},
\label{tanhw}
\end{equation}
in which $\mu$ can only take value from $\{1,2,3\}$. So
$w^{\mu}_{ij}$ is one of the three relationships---coauthor,
citation or acknowledgement and is defined as
\begin{equation}
w^{\mu}_{ij}=\tanh\left(\alpha_{\mu}T^{\mu}_{ij}\right),
\label{coauthor}
\end{equation}
where $T^{\mu}_{ij}$ is the time of $\mu$ relationship between $i$
and $j$.

As we mentioned in introduction section, we think the weight
should not increase linearly, and it must reach a limitation when
the time exceeds some value. So we use tanh function to describe
this nonlinear effect. We also assume the contributions to the
weight from these three relations are different and they can be
represented by the different values of $\alpha_{\mu}$.
$0.7,0.2,0.1$ are used for $\alpha_{1}, \alpha_{2}, \alpha_{3}$ in
this paper. The effect of different coefficient could not be
revealed by any of the quantities analyzed so far.

The similarity is used here as the weight, after the network has
been constructed, it is converted into dissimilarity weight as
\begin{equation}
\tilde{w}_{ij} = \frac{3}{w_{ij}} \hspace{0.3cm}( if
\hspace{0.15cm} w_{ij} \neq 0 ).
\end{equation}
It's timed by $3$ because the similarity weight
$w_{ij}\in\left[0,3\right]$. Therefore, we have
$\tilde{w}_{ij}\in\left[1,\infty\right]$, and it is corresponding
to the "distance" between nodes. All quantities are calculated
under this dissimilarity weight from now on if not mentioned.

\begin{figure}
  \includegraphics[width=6cm]{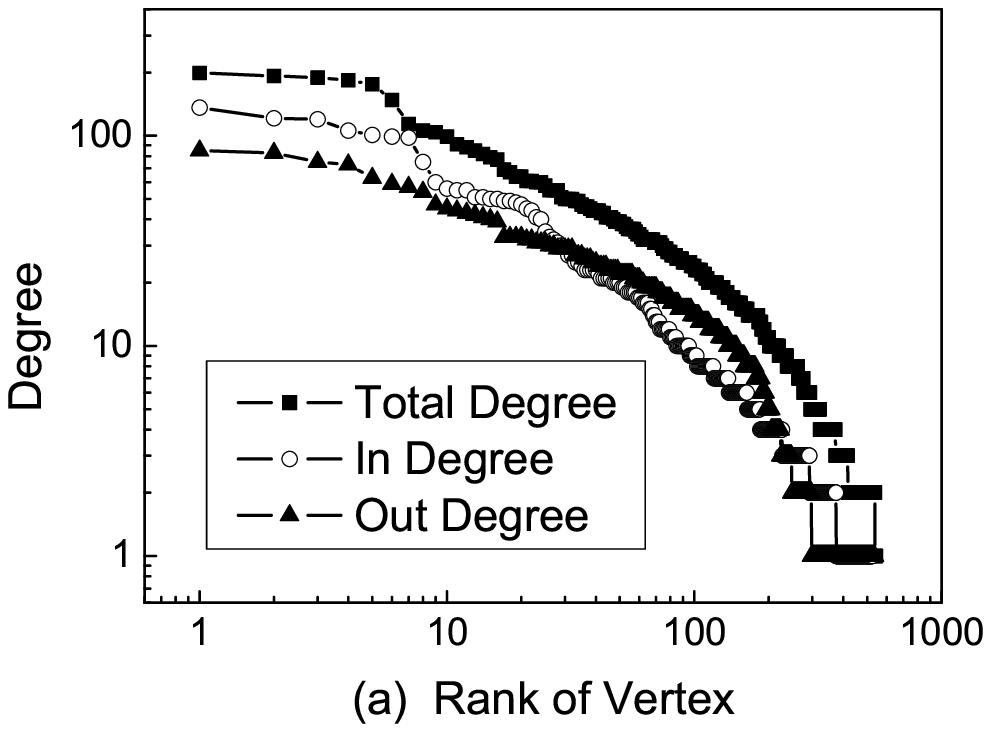}\includegraphics[width=6cm]{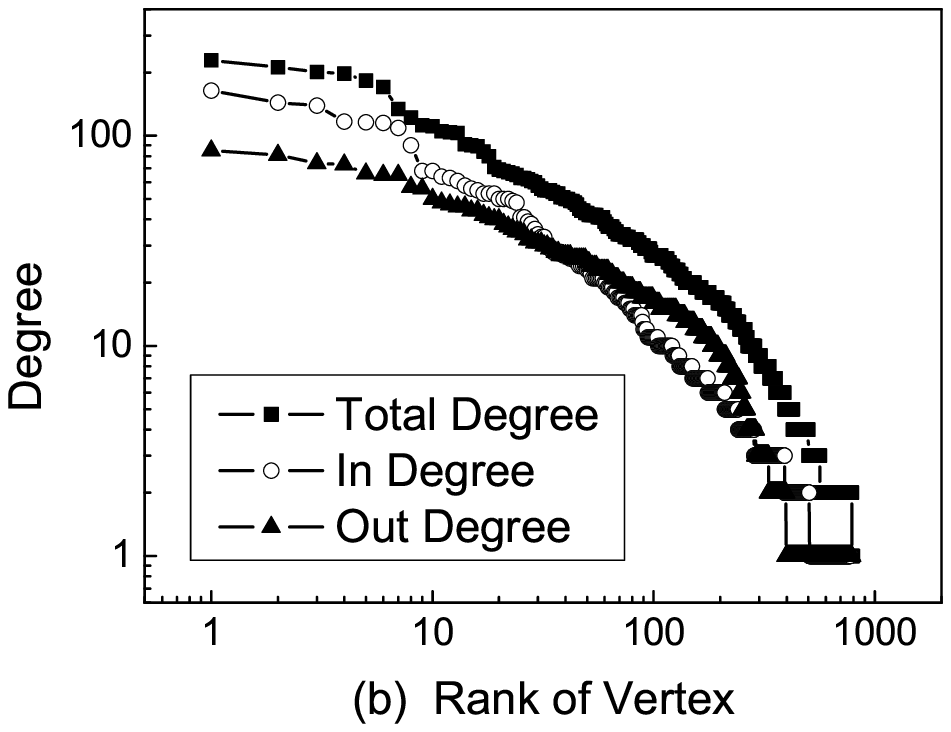}\\
  \includegraphics[width=6cm]{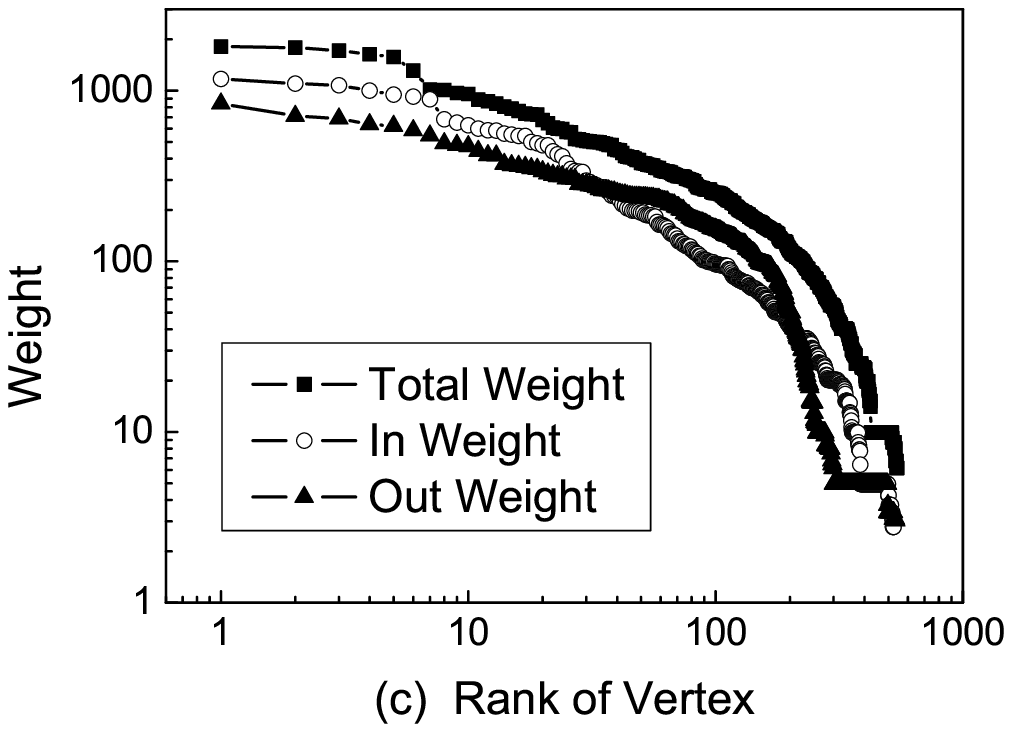}\includegraphics[width=6cm]{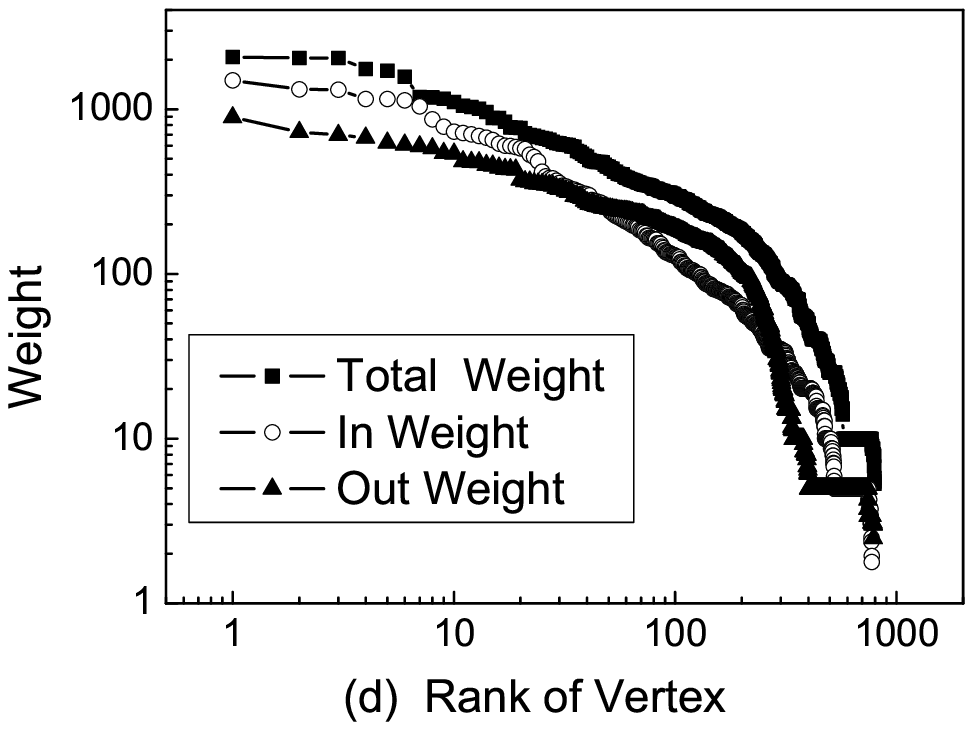}
  \caption{Zipf plot of degree and weight for different data set.
Degree distribution for 2003(a) and 2004(b). Weight distribution
for 2003(c) and 2004(d).}\label{comD&W}
\end{figure}

\begin{figure}
  \includegraphics[width=6cm]{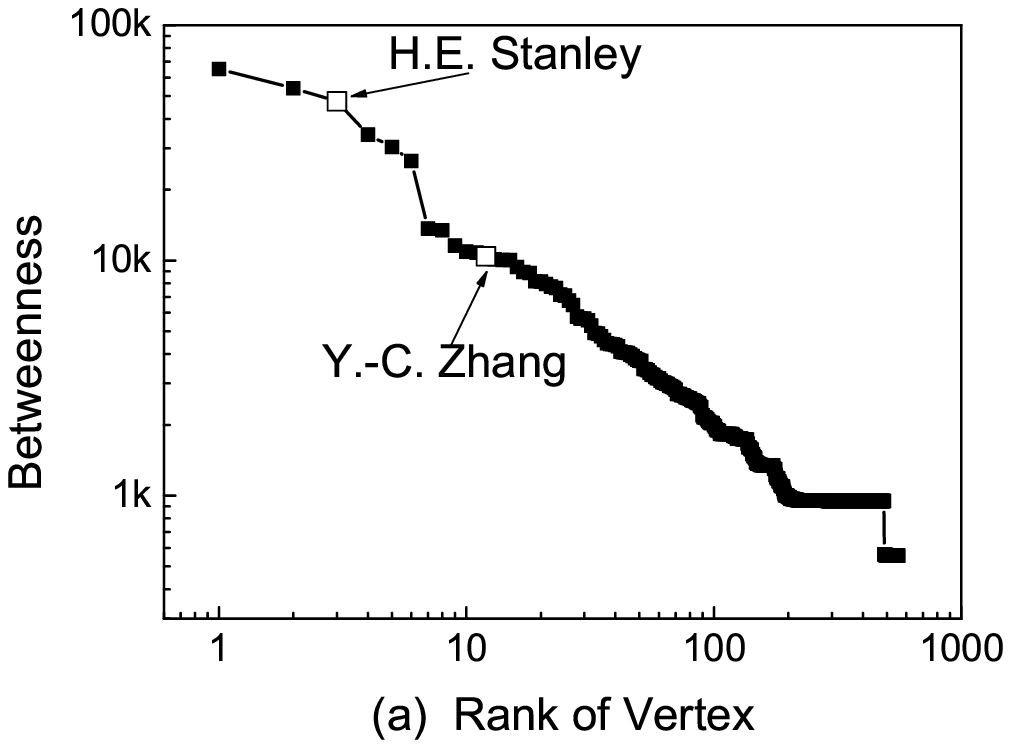}\includegraphics[width=6cm]{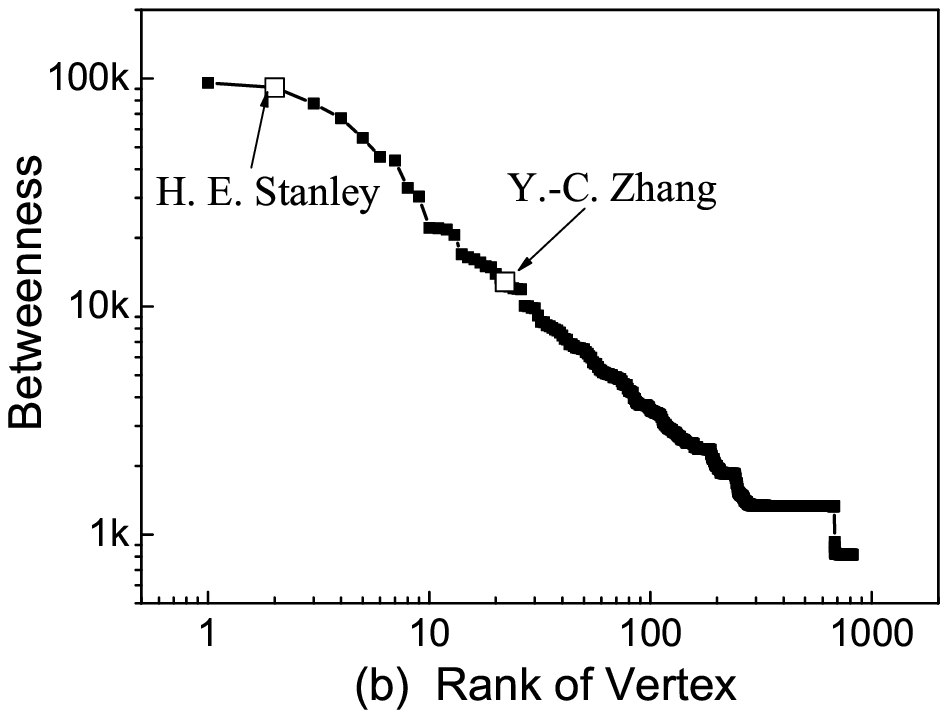}\\
  \caption{Zipf plot of vertex betweenness for 2003(a) and 2004(b).}\label{comVB}
\end{figure}

It is interesting to compare the basic statistical results of the
enlarged data set with the results given in paper \cite{fan}.
Fig.\ref{comD&W} gives the results for degree and weight
distribution in Zipf plots. The qualitative properties are
unchanged, but detailed structures such as the position of a
certain vertex have been changed. Fig.\ref{comVB} are the vertex
betweenness for two data sets. Although the qualitative properties
are the same, the position of vertex has been changed. We label
the positions of Prof. H. E. Stanley and Prof. Y.-C. Zhang as
examples. In the development of Econophysics, Stanley is
well-known by a series of pathbreaking works on empirical and
modelling analysis of time series of economical data, such as
stock prices and firm sizes, and Zhang contributed the a
significant step in Minority Game, an easy-understood but fruitful
model for collective decision making in economic world. These
changes may reflect the development of Econophysics from the view
point of network analysis. For example, we can choose a group of
people working on one aspect and then tracing their position in
the plot given above. It's easy to know the whole picture of the
recent development of this group relative to the others.

\begin{figure}
  \includegraphics[width=12cm]{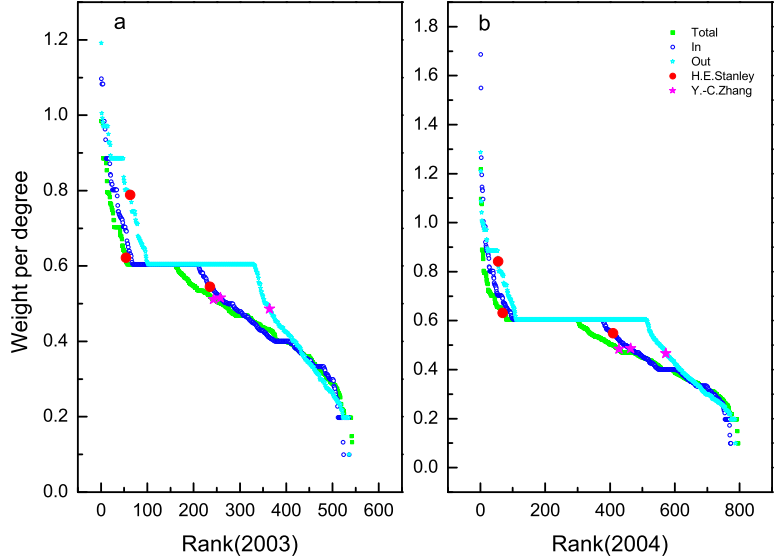}\\
  \caption{Zipf plot of total, in and out weight per degree for 2003(a) and 2004(b).
  The points are marked are the WPD values of H.E.Stanley and Y.-C. Zhang. The
  platforms in all the curves suggest the working style of a large group of vertex. Weight on the
  platforms is about $0.604$, which is roughly $tanh\left(0.7*1\right)$. This looks
  like those people are connected to the community just by one cooperation.}\label{wpd}
\end{figure}

In \cite{fan}, we also introduced weight per degree (WPD), a
characteristic quantity of vertex. In \cite{fan}, it was defined
by total weight divided by total degree of every vertex. Now we
want to present more detail of this quantities by out-WPD, in-WPD
and total-WPD. Out-WPD is the quotient between the strength of
outgoing relationship and the number of outgoing edges, so this
represents how actively the vertex communicate with others, more
intensively or more extensively. And the in-WPD represents how
intensively or extensively the community treated the specific
vertex. For a pioneer scientist in a field, the vertex will have
more edges other than more weight (times) on edges, while an
evergreen vertex probably will have more weight (times) on edges
other than more edges. So WPD provides a character of the working
style of the vertex. For example, from fig\ref{wpd}, we can see
the out-WPD of Stanley is quite large compared with the other two
WPDs of him, or even compared with the out-WPD of Y.-C.Zhang. In
some senses, this implies Stanley is a little bit more outgoing
than Zhang, as the figure suggested. Any way, here we proposed
this quantity only for fun, however, we wish later on it will be
found some good meaning in reality such as Social Network
Analysis, hopefully.

\section{Clustering Coefficient and the Role of Weight}
Now we turn to the effects of weight on the structure of weighted
network. First, we introduce the way to varying the relation
between weights and edges to investigate the role of weight on the
structure. And then, compare the different behaviors of
topological quantity to reveal the effect of such variation.
Especially, the clustering coefficient of weighted networks will
be defined and discussed. And then, the average shortest path and
betweenness will be calculated and compared.

\subsection{Clustering and Distance}
\label{secweight}

It is well-known that the efficiency of small world network and
scale-free network in real world is characterized by the
coexistence of small relative distance
$\frac{L\left(p\right)}{L\left(0\right)}$ and high relative
clustering coefficient $\frac{C\left(p\right)}{C\left(0\right)}$
compared with the distance $L\left(0\right)$ and cluster
coefficient $C\left(0\right)$ of the induced regular network with
the same number of vertex and edges. For a weighted network, a new
type of random network can be introduced. The weights on edges can
be randomized in weighted networks, while in non-weight networks,
the only thing can be randomized is the link. This effect on the
network structure is new in weighted networks, and it can be
interpreted as the topological role of weight, which tells us
whether the weights are distributed randomly or are related with
the inherent structure.

The general approach is to change the relationship between weight
and edge at a specific level $p$. Set $p=1$ represents the
original weighted network given by the ordered series of weights
which gives the relation between weight and edge but in a
decreasing order,
\begin{equation}
W(p=1) = \left(w_{i_{1}j_{1}} = w^{1}\geq w_{i_{2}j_{2}} =
w^{2}\geq \cdots \geq
w_{\left(i_{L}\right)\left(j_{L}\right)}=w^{L}\right).
\end{equation}
$p=-1$ is defined as the inverse order as
\begin{equation}
W(p=-1) =\left(w_{i_{1}j_{1}} = w^{L}\leq \cdots \leq
w_{\left(i_{L-1}\right)\left(j_{L-1}\right)} = w^{2} \leq
w_{\left(i_{L}\right)\left(j_{L}\right)}=w^{1}\right),
\end{equation}
which assign the minimum weight to the edge with a maximum weight
in the origin network, and so on. And $p=0$ represents a fully
randomized series of $W(p=1)$.
\begin{equation}
W(p=0) = FullyRandomized\left(w^{1}, w^{2}, \cdots, w^{L}\right).
\end{equation}
Therefore, $p$ in some senses behaviors like a correlation
coefficient between the new and the origin weight series. If we
know how to generate a random series from a given series with
fixed correlation $p$, then we can plot all the relative cluster
coefficients and relative distance vs $p$ just like the famous
figure in \cite{swn}. The way to generate a conditional random
series from a given series is so-called ``conditional uniform
graph tests''\cite{test}, which has more general sampling
procedures to randomize a given series. However, in this paper, we
only investigate the special cases corresponding to $p=1,0,-1$.
The induced fully randomized weighted network is constructed by
keeping the ordered set of edges but randomizing choosing values
from the set of weights. Every edge is given a weight randomly
selected from the weight set. Then we compared the basic
topological properties of the original networks with the inverse
or randomized one.

For a directed network, the nearest neighbor of a vertex can be
defined as In, Out and Total, so the clustering coefficient of a
directed network also has these three different quantities, named
as $Icc$, $Occ$ and $Tcc$ for short. Let's take $Icc$ for
instance. For every vertex $v_{i}$ in the network, the vertex
having edge ending at $v_{i}$ forms a neighbor set $\partial_{i}$
of vertex $v_{i}$. Then $Icc$ is defined as
\begin{equation}
Icc = \left\langle Icc_{i}\right\rangle_{i} = \left \langle
\frac{M_{i}}{|\partial_{i}|(|\partial_{i}|-1)} \right \rangle_{i}.
\end{equation}
where
\begin{equation}
M_{i}=\sum_{j,k\in\partial_{i}}\frac{1}{\tilde{w}_{jk}},
\end{equation}

This definition will give the same value as the usual clustering
coefficient for directed non-weighted networks, and half of the
corresponding value for undirected and non-weighted networks. The
meaning of the numerator is the summation of all similarity among
the neighborhood, while the denominator is the possible maximum
value of similarity among them and the maximum value can be
reached if everyone of the neighborhood is connected to each other
and all the value of similarity is $1$. So this definition has the
same meaning of the clustering coefficient of non-weighted network
but take the weight of links into account.

The average shortest distance $d$ is defined as,
\begin{equation}
d = \frac{1}{N(N-1)}\sum_{ij} d_{ij} \label{avedistance}
\end{equation}\label{distance}
in which $d_{ij}$ is the shortest distance between vertex $i,j$
and equals to $N$ (the size of network) if no path exists.

The above definition is used to calculate the clustering
coefficient and average shortest distance for weighted networks.
Table 1 gives the clustering coefficients for the real, inverse,
and randomized network respectively constructed by the data set of
Econophysists and the data set of scientists collaboration
provided by Newman. The later data set has the only times of
collaboration between scientists, so the corresponding network is
a weighted but not a directed one. The weight is given by the
measurement we introduced in last section. For the fully
randomized network, the result of clustering coefficient is the
average of 100 random samples. In next section, the results of
distribution of betweenness for fully randomized networks are also
the average of 100 sampling processes. It is interesting to find
that the clustering coefficient for the real network of
Econophysists is obviously larger than the inverse and randomized
one. It's similar with situation of WS small world\cite{swn},
where the randomization also leads to a gradual decrease of
cluster coefficient. The difference on clustering coefficient
among real, inverse and randomized network also implies that there
is certain relationship between weight and inherent network
structure.

\begin{center}
Table 1 Clustering coefficients of weighted network
\end{center}
\begin{center}
\begin{tabular}{|c|c|c|c|c|}
\hline
\multicolumn{2}{|c|}{} & Real & Inverse & Random   \\ 
\hline
 & Tcc & 0.064 & 0.029 & 0.038  \\ \cline{2-5}
\raisebox{1.7ex}[0pt]{Clustering}& Icc& 0.057 & 0.033 & 0.037  \\
\cline{2-5}
\raisebox{1.7ex}[0pt]{coefficients}& Occ & 0.067 & 0.015& 0.029 \\
\hline
Newman\cite{datanote}& Tcc & 0.430 & 0.407& 0.400 \\
\hline
\end{tabular} \label{cluster}
\end{center}
There seem to be a large difference on clustering coefficient
between our network and Newman's collaboration network. The most
significant reason is that the largest connected cluster dominant
almost perfectly in Newman's data (83.2 percent of nodes are in
the largest cluster), while in our case, only $25.3\%$ nodes are
in the largest cluster when the co-authorship is considered. The
link in our network are dominated by the directed links for
citations with larger weight of dissimilarity. The mean value of
the weight per degree of the two network confirms such an
argument: $\left(2.60, 1.52\right)$ in dissimilarity or
$\left(0.48, 0.68\right)$ in similarity for our network and
Newman's separately. This means the average length of our edge is
much longer than Newman's, or we say, the relationship in Newman's
network is much strong than ours. This is easy to understand
because the topics covered in Newman's data is much better
developed than the topics covered by our network. If we compare
our clustering coefficient from largest connect cluster, $0.093$,
and it should times a factor of $2$ because of the difference of
directed and non-directed network, then it comes to $0.186$.
Considering further the ratio of the weight per degree of the two
network, that is $2.60/1.52=1.7$, the coefficient comes to
$0.316$. This is still smaller than $0.430$, the result of
Newman's network, yet comparable. Of course, it's true that there
is a long way to go to develop Econophysics into a similar
developed stage of Physics.

From the definition of average shortest distance expressed by
formula \ref{avedistance}, for the sparse network, the average
shortest distance is dominated by the isolated vertices or small
clusters, because the distance between any two disconnected vertex
is set as $N$ the size of the network. In Table 2, we give only
the corresponding results for the largest connected cluster. The
average shortest distance is the result of corresponding
undirected cluster(if there are two directed edges between two
nodes, we simply dropped the edge with smaller weight). The
weight-randomized network has also smaller average shortest
distance and clustering coefficient. Again, this implies weight
has some inherent relation with structural role of edge.

\begin{center}
Table 2 Results for the largest cluster
\end{center}
\begin{center}
\begin{tabular}{|c|c|c|c|c|}
\hline
 & \multicolumn{3}{|c|}{weighted} &  \\   \cline{2-4}
  & Real & Inverse & Random & \raisebox{1.5ex}[0pt]{Non-weighted} \\ \hline
 Tcc & 0.093 & 0.050 & 0.067 &0.363 \\ \hline
  d & 22.91 & 21.83 & 17.75 &3.217 \\ \hline
 \end{tabular}\label{distan}
\end{center}

In the right column of Table 2, the corresponding results for
non-weighted cluster are given. We can not compare these values
with that of weighted networks. So we have compared the
distribution of link and vertex betweenness for weighted and
non-weighted cluster in Fig. \ref{wnonw}. We find that the weight
affects the distribution, but leads to qualitatively similar
results. However, the detail according to every single vertex is
different. This is shown by the two small hollow rectangles
representing the same person on the two curves. Although in the
above studies we have found some effects of weight distributions,
it seems that other quantities and their distributions may be
needed to investigate the topological role of weight.

\subsection{Distribution of Clustering Coefficient and Betweenness}
\begin{figure}
  \includegraphics[width=6cm]{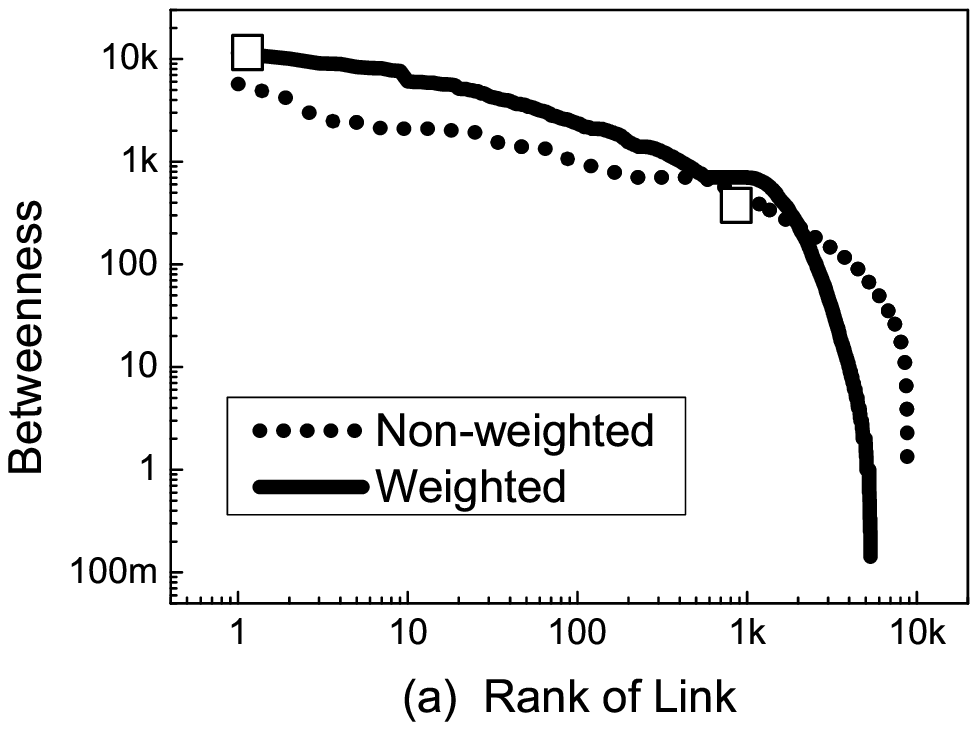}\includegraphics[width=6cm]{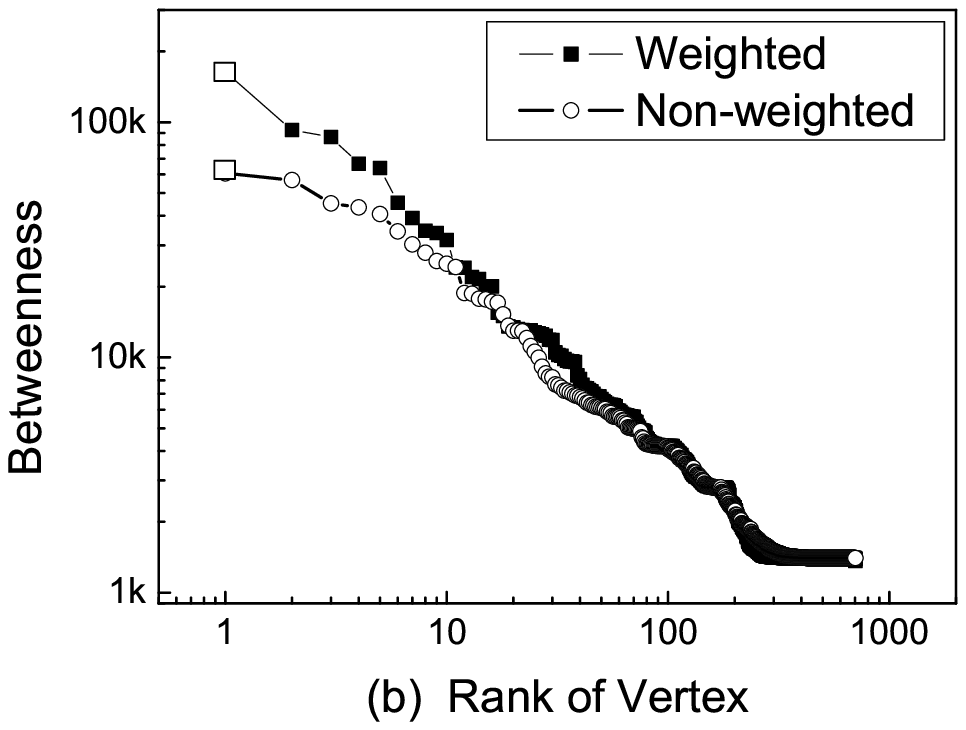}\\
  \caption{Comparison of weighted and non-weighted largest
 cluster.(a) Link betweenness, (b) Vertex betweenness. }\label{wnonw}
\end{figure}
\begin{figure}
  \includegraphics[width=6cm]{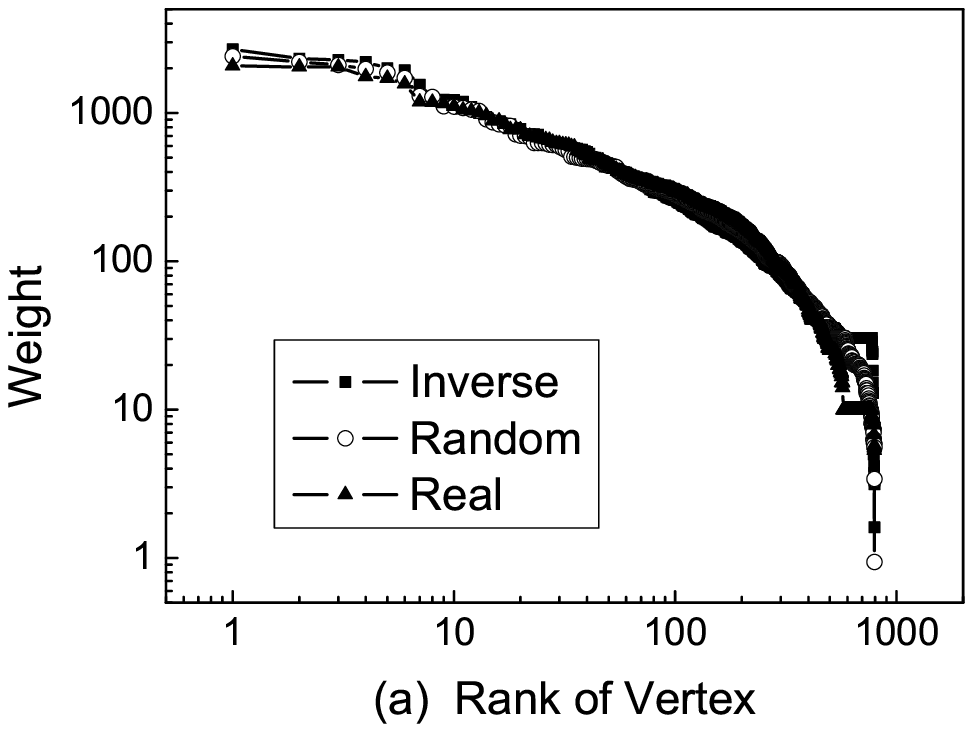}\includegraphics[width=6cm]{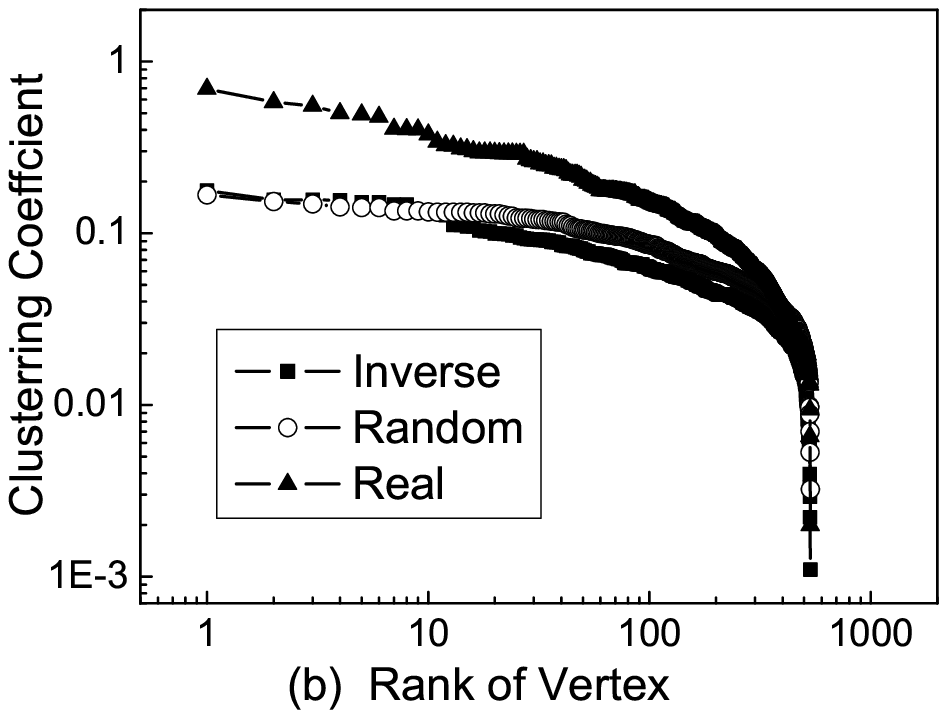}\\
  \includegraphics[width=6cm]{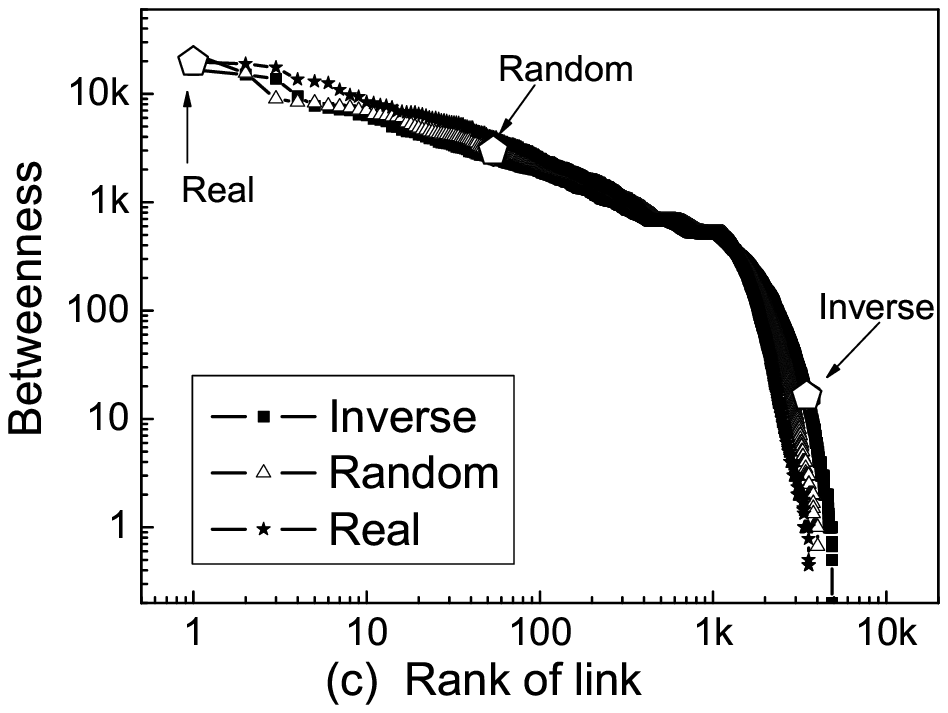}\includegraphics[width=6cm]{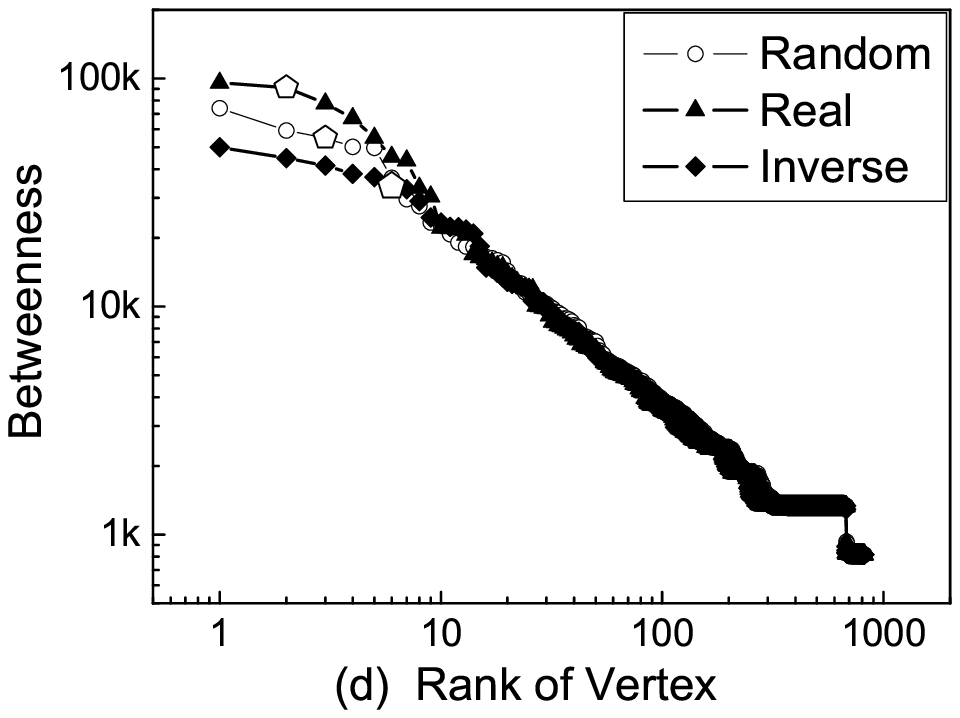}\\
  \caption{Comparison of real, inverse, and randomized weighted
networks. (a) Vertex weight distribution, (b) Clustering
coefficient, (c) Link betweenness, (d) Vertex betweenness. All in
Zipf plots.}\label{VLbetween}
\end{figure}

In order to study the impaction of weight to the topological
properties of network, we have introduced the way to re-assign
weights onto edge with $p=1,0,-1$ for weighted networks. Besides
the average clustering coefficient and average shortest distance,
the change of distribution of corresponding topological quantities
should give more detailed descriptions for the effects of weight.
Fig. \ref{VLbetween}(a) and (b) give the weight and clustering
coefficient distribution for real, inverse, and randomized
weighted network. It seems that in all the cases the vertex weight
distribution keeps the same while the distribution of clustering
coefficient changes obviously.

Other important global and structural quantities of a network to
investigate the impaction of weight on the structure are the
vertex betweenness and link betweenness. Fig.\ref{VLbetween}(c)
and (d) give the distribution of the vertex betweenness and link
betweenness in Zipf plot for all cases. The distribution of link
betweenness seems unchanged but if we focus on the position of a
certain edge (the top one in real weighted network for instance)
in the curves, it changes a lot for different cases. From the
comparison here and in Fig.\ref{wnonw}, we know the way of weight
measurement affects the structure of network, but not described
very well by quantities all above. We assume correlation analysis
will provide more detailed information beyond this, because the
measurement of weight has different effects on different
quantities. For example, it doesn't affect the degree of vertex at
all, but does affect betweenness of vertex. Maybe a correlation
analysis between such quantities will tell more about the
character of weighted networks. However, in
Fig.\ref{VLbetween}(d), the upper tail of the distribution for the
vertex betweenness has been changed, but the position of a vertex
does not change as much as links. It seems that the betweenness of
vertex is dominated by links more than by the weights on the
links.

\section{Conclusions}
From the comparison between networks with real, inverse,
randomized weight and induced non-weight networks, we know the
network structure depend on the weight. We calculated global
structure quantities as clustering coefficient, betweenness of
vertex and betweenness of edge under different cases. Even some
global distribution seems robust but the detailed structure has
been affected by the weight. These results give us clues to the
question of the topological role of weight. As we point out in
section \ref{secweight}, this question investigates the
relationship between weight and inherent network structure. It
sounds like strong correlation existing between them. But it seems
that other quantities and distributions are needed to investigate
the topological role of weight. The conclusion depends on the more
general exploration in more networks and modelling research.

In summary, in this paper, we have constructed a small network by
collecting papers in Econophysics. A new definition of weight and
new topological properties are introduced and some fundamental
properties are analyzed, including preliminary analysis of the
topological role of weight. The idea to integrate networks with
multilevel but the same kind of relationship have further
application value. We wish more data can be collected including
the time development of the network so that it will help to
analyze the evolution of networks, especially for the networks of
scientist, in which the network structure and the dynamical
phenomena such as exchanging idea are in co-evolution. In this
sense, network of idea transportation has some special value,
because the network structure behind the dynamical behavior over
the network are always coupled together. So dynamical process over
the network and the evolution of this network are in fact always
entangled each other. As in the subway analogy, we want to extract
the information about railway from traffic, but at the same time,
in our network of scientists, the traffic can generate new paths!

Therefore, works on modelling such network will have very
important and special value. Inspired by the empirical study in
this paper, recently we have proposed a model of weighted network
showing almost exactly the same behavior qualitatively. The most
important character of the model is that the only dynamical
variable is the times of connection, not two variables as both of
connection and weight as in usual models of weighted networks.
Hopefully, in the near future, we can complete the modelling work
and compare the results with the empirical results here.

\section{Acknowledgement}
The authors want to thank Dr. Newman for his cooperation data,
thank Dr. Yougui Wang and all other group members for the
inspiring and warm discussion. This project is supported by NSFC
under the grant No.70471080 and No.70371072.


\begin{thebibliography}{99}
\bibitem{datanote}In this calculation, we use the data from Newman, a data set of scientists with a much larger size. In this network, the Cluster Coefficient is much larger than the one of our data. Probably, bacause Econophysics is not as well developed now as traditional topics.
\end{thebibliography}
\end{document}